\documentclass[10pt]{article}
\usepackage[margin=1in]{geometry}

\usepackage[affil-it]{authblk} 
\usepackage{etoolbox}
\usepackage{lmodern}
\usepackage{rotating}
\makeatletter
\patchcmd{\@maketitle}{\LARGE \@title}{\fontsize{14}{17.2}\selectfont\@title}{}{}
\makeatother

\usepackage{sectsty}

\sectionfont{\fontsize{13}{15}\selectfont}
\subsectionfont{\fontsize{12}{15}\selectfont}

\usepackage[font=footnotesize]{caption}
\usepackage{authblk}
\usepackage{microtype}
\usepackage{graphicx}
\usepackage{subfigure}
\usepackage{booktabs} 
\usepackage{amsmath,amssymb, amsthm, bm}
\usepackage{color,soul}

\usepackage{setspace}
\usepackage{gensymb}
\usepackage{bbm}
\usepackage[colorlinks=true,allcolors=blue]{hyperref}
\usepackage{natbib}
\usepackage{url}
\usepackage{wrapfig}
\usepackage{comment}
\usepackage{amsmath,amsfonts,amssymb}

\usepackage{mathtools}

\usepackage[ruled,vlined]{algorithm2e}
\usepackage{amsmath}
\usepackage[makeroom]{cancel}

\DeclarePairedDelimiterX{\infdivx}[2]{(}{)}{%
  #1\;\delimsize\|\;#2%
}

\newcommand{\norm}[1]{\left\lVert#1\right\rVert}

\newtheorem{assumption}{Assumption}

\newtheorem{subassumption}{Assumption\normalfont}[assumption]
\newenvironment{assumption*}
 {\ifnum\value{subassumption}=0 \stepcounter{assumption}\fi\subassumption}
 {\endsubassumption}
\newenvironment{assumption+}[1]
 {\renewcommand{\thesubassumption}{#1}\subassumption}
 {\endsubassumption}

\newtheorem{lemma}{Lemma}

\newtheorem{exmp}{Example}[section]
\usepackage{colortbl}

\usepackage{amsmath}

\DeclareMathOperator*{\argmin}{arg\,min}

\DeclareMathOperator{\btheta}{\bm{\theta}}
\DeclareMathOperator{\bbtheta}{\bm{\bar{\theta}}}

\usepackage{fontawesome5}
\usepackage{hyperref}
\makeatletter
\newcommand{\github}[1]{%
   \href{#1}{\faGithubSquare}%
}
\makeatother

\title{\textbf{Federated Data Analytics: A Study on Linear Models}}
\date{\vspace{-7ex}}

\author[1]{Xubo Yue}
\author[1]{Raed Al Kontar}
\author[2]{Ana Mar\'ia Estrada G\'omez}
\affil[1]{Industrial \& Operations Engineering, University of Michigan, Ann Arbor}
\affil[2]{School of Industrial Engineering, Purdue University, West Lafayette}

\begin{document}
\maketitle

\begin{abstract}
As edge devices become increasingly powerful, data analytics are gradually moving from a centralized to a decentralized regime where edge compute resources are exploited to process more of the data locally. This regime of analytics is coined as federated data analytics (FDA).  In spite of the recent success stories of FDA, most literature focuses exclusively on deep neural networks. In this work, we take a step back to develop an FDA treatment for one of the most fundamental statistical models: linear regression. Our treatment is built upon hierarchical modeling that allows borrowing strength across multiple groups. To this end, we propose two federated hierarchical model structures that provide a shared representation across devices to facilitate information sharing. Notably, our proposed frameworks are capable of providing uncertainty quantification, variable selection, hypothesis testing and fast adaptation to new unseen data. We validate our methods on a range of real-life applications including condition monitoring for aircraft engines. The results show that our FDA treatment for linear models can serve as a competing benchmark model for future development of federated algorithms.
\end{abstract}

\section{Introduction}
\label{sec:intro}

The sheer amount of data collected nowadays is beginning to overwhelm traditional centralized data analytics regimes where data from the edge is continuously uploaded to a central server to be processed. Excessive communication traffic from data upload, significant central server storage needs, energy expenditures from centralized learning of big data models, and privacy concerns from sharing raw data are becoming critical challenges in centralized systems. Statista predicted that, by 2024, data produced on edge devices (e.g., cell phone data, self-driving vehicle data) would reach more than hundreds of zettabytes while the global central servers only have 10.4 zettabytes of storage \citep{morell2022dynamic}. Transmitting such a vast amount of edge data into a central server is infeasible. Adding to that, training a model with moderately large datasets results in significant budget costs and carbon emissions \citep{patterson2021carbon}. Furthermore, data-sharing comes with serious privacy concerns. According to \cite{lawson2015connected}, Canadian drivers who refused to enroll in the automotive telematics program demanded that their personal driving data (e.g., behavior, location, web-browsing history) should be respected by vehicle companies and that they be given control over the data collection process. These debates over data protection standards have not faded away over the past decade.

Fortunately, the Internet of Things (IoT) is undergoing a new revolution in which the compute power of edge devices  is tremendously increasing \citep{hassan2018role}. AI Chips such as general-purpose chips (GPUs), semi-customized chips (FGPAs) and fully-customized chips (ASICs) are becoming readily available across many applications \citep{blanco2019deep, rahman2021internet, zhu2021green}. Such AI chips are able to process a vast amount of data locally and provide timely responses and decisions \citep{shi2016edge}.  For instance, the autonomous vehicle company PerceptIn has released a real-time edge computing system, DragonFly+, that is three times more power efficient and delivers three to five times of the computing power of a Nvidia Tx1 and an Intel Core i7 processor \citep{liu2019edge}. Another notable example is Tesla's autopilot system that has computing power on the car itself comparable to hundreds of MacBook pros \citep{Tesla}. As a consequence, traditional IoT is on the verge of shifting to a decentralized framework recently termed the Internet of Federated Things (IoFT) \citep{kontar2021internet} in which some of the data processing is deferred to the edge. In this future, the central server only acts as an orchestrator of the learning process and an integration point of model updates from different devices, rather than the central location where all data is processed. Indeed, IoFT is slowly infiltrating various fields such as manufacturing, transportation, and energy systems \citep{kontar2021internet}. 

The underlying data analytics framework in IoFT is federated data analytics (FDA) where edge devices exploit their own computation power to collaboratively extract knowledge and build smart analytics while keeping their personal data stored locally. Consequently, edge devices no longer need to upload their data to the cloud (or server), and, in turn, the cloud does not need to store that immense amount of data. As such, FDA resolves many of the aforementioned drawbacks of the centralized computing system and sets forth many intrinsic advantages including privacy-preserving and reducing storage/computation/communication costs, among many others.

In spite of some recent advances in FDA, most, if not all, literature focuses on deep neural networks \citep{li2020federated1} (learned via first order methods). To date, very few papers have delivered a federated treatment of traditional statistical models. Perhaps the closest field where statistical models were investigated is distributed learning (DL) \citep{jordan2018communication}, yet DL and FDA have several fundamental differences. Despite the terminology ``distributed", DL is still a centralized computation approach where different compute nodes operate on all data \citep{fan2021communication}. These nodes communicate often, observe each others data and can operate on different data partitions. The underlying philosophy for DL is ``divide-and-conquer'' where data is divided across the nodes (often dynamically), and then the nodes collaborate to ``conquer" (learn) a single model. In constrat, in IoFT data resides at the edge and cannot be shuffled, randomized, and divided. Therefore, edge devices cannot see each others data and data partitions for FDA are fixed and often heterogeneous.  Besides that, devices have datasets with unique features as they correspond to different clients, components or systems (e.g., cars). As such, in FDA there is often no single model to conquer, rather, our goal is to borrow strength across edge devices to improve our analytics.

In this work, we take a step back and move out of the regime of deep neural networks to study one of the most fundamental statistical models: linear regression (LR). Indeed, linear models may facilitate hypothesis testing, uncertainty quantification, variable selection, deriving engineering insight and establishing a baseline to compare other models with. Needless to say, in reality many real-applications can be sufficiently characterized by linear models \citep{liu2013data, si2017distribution, li2018monitoring, schulz2020different,arashi2021ridge, csahin2022linregdroid}. In addition, building upon FDA for linear models, one may develop approaches for more complex derivatives such as  logistic regression, mixed-effects and kernel methods. 

To this end, we exploit the properties and structure of linear models and develop an FDA treatment for linear regression with Gaussian noise, entitled \texttt{FedLin}. Our treatment is built upon hierarchical models (HM) which allow borrowing statistical knowledge across groups (i.e., devices or clients in FDA). Specifically we propose two federated HM structures that provide a shared representation across devices to facilitate information transfer. The first structure establishes a shared representation defined through a structural prior over concatenated device parameters. The second structure is based on the assumption that all device parameters are generated from the same underlying distribution. This allows uncertainty quantification and consequently a Bayesian treatment for variable selection in \texttt{FedLin}. Our methods are validated on a range of real-life problems including variable selection and condition monitoring. The results highlight the effective performance of our approaches and their ease of implementation which may help them serve as  benchmark models for many future developments of federated statistical algorithms.

\textbf{Organization:} The remainder of this paper is organized as follows. In Section \ref{sec:general}, we conduct a literature review and introduce the general setting and motivation. In Sections \ref{sec:HM1} and \ref{sec:HM2}, we present our two model structures and their applications. We validate our proposed models on various simulated and real-life datasets in Sections \ref{sec:sim} and \ref{sec:exp}. Finally, we conclude our paper in Section \ref{sec:con}. Besides, codes for this paper, written in R language, can be found at \url{https://github.com/UMDataScienceLab/Federated-Linear-Models}.
\section{Background}
\label{sec:general}

\subsection{Literature Overview}
\label{sec:related}

The idea of FDA was first brought to the forefront of deep learning by \cite{mcmahan2017communication}. In this work, they proposed the FDA algorithm termed federated averaging (\texttt{FedAvg}). The idea of \texttt{FedAvg} is simple: a central server distributes initial deep learning model parameters and the network structure to some selected devices, devices perform local stochastic gradient descent (SGD) steps using their data and send their updated parameters back. The server then takes an average of those parameters to update the global model. This process is termed as one communication round and is iterated several times. Although simple, \texttt{FedAvg} is still one of the most competitive benchmark models nowadays \citep{kairouz2021advances}. To date, some work has been proposed to improve the performance of federated deep learning algorithms. For instance, \cite{yuan2020federated} and \cite{liu2020accelerating} provided several provable techniques to accelerate \texttt{FedAvg} and enable faster convergence. \cite{li2019fair,yu2020fairness,yue2021gifair,du2021fairness} developed variants of  \texttt{FedAvg} that ensure uniformly good performance across all devices to achieve fairness. Another line of work aims to develop personalized solutions in federated data analytics as excessive heterogeneity can greatly impact the performance of a single global model \citep{deng2020adaptive, fallah2020personalized,li2021ditto}. Such approaches usually either follow a train-then-personalize philosophy where a trained global model is fine-tuned on local devices or divide the layers of a neural network into shared and individualized ones \citep{tan2022towards}, where devices collaborate to learn the common layers using methods such as \texttt{FedAvg}.  From a theoretical perspective, \cite{stich2018local, li2020federated} prove the convergence of \texttt{FedAvg} for convex functions and homogeneous (\textit{i.i.d.}) datasets. Those results are then extended to a non-convex setting by \citet{wang2021cooperative}. On the other hand, \cite{li2019convergence} extend the results of \cite{stich2018local} to the non-\textit{i.i.d.} setting. Furthermore, \cite{shi2021fed} extend the convergence results to a kernel regime. For a comprehensive overview of current literature, please refer to \cite{kontar2021internet}.

The major trend of FDA exclusively focuses on neural networks and classification tasks. FDA for statistical models is still scant. \cite{yue2021federated} extend the Gaussian process to a federated framework and show that their proposed algorithm can achieve state-of-the-art performance on multi-fidelity modeling problems. \cite{yuan2021federated} develop a federated composite optimization framework that solves the federated lasso problem. \cite{tong2020federated} propose a federated iterative hard thresholding algorithm to tackle the non-convex $0$-norm penalized regression problem. The two aforementioned papers mainly formulate penalized regression from a frequentist perspective. In Sec. \ref{sec:HM2}, we will develop a Bayesian formulation build upon HM for federated penalized regression.


\subsection{General Setting}
\label{sec:general-1}

We start by describing our problem setting.  Suppose there exists $K\geq 2$ edge devices. For device $k\in[K]\coloneqq\{1,\ldots,K\}$, the dataset is given as $\bm{D}_k=\{\bm{X}_k,\bm{Y}_k\}$ with $N_k$ observations, where $\bm{Y}_k=[y_{k1},\ldots,y_{kN_k}]^\intercal$ is a $N_k\times 1$ output vector, $\bm{X}_k=[\bm{x}_{k1},\ldots,\bm{x}_{kN_k}]$ is a $d\times N_k$ input matrix and $\bm{x}_{k1}=([\bm{x}_{k1}]_1,\ldots,[\bm{x}_{k1}]_d)^\intercal$. Here, $d$ is the dimensionality of the input space. In this work, we focus on linear models. More specifically, data on device $k$ is used to learn a linear model parameterized by $\btheta_k\in\mathbb{R}^d$. The distribution of $y_{ki}$ is given as
\begin{align}
\label{eq:y}
    y_{ki}|\bm{x}_{ki},\btheta_k\sim\mathcal{N}(x_{ki}^\intercal\btheta_k,\sigma_k^2), \quad \forall i=1,\ldots,N_k,
\end{align}
where $\sigma_k^2$ is a noise parameter. For the sake of compactness, denote by $\bm{\Theta}=(\btheta_1,\ldots,\btheta_K)$ a $d\times K$ matrix concatenating all device parameters.  

Further, we assume that a central server is connected to all edge devices and can facilitate the collaborative model learning process. As such, our goal in FDA is for devices to leverage their commonalities to better learn model parameters $\bm{\Theta}$; all while distributing the learning efforts and circumventing the need to share raw data.


\subsection{Federated Data Analytics and Hierarchical Models}
\label{sec:general-2}

Since our goal is to borrow strength across devices, the first step is to create a shared representation across individual device models in order to facilitate the inductive transfer of knowledge. Here we adopt the natural hierarchy in FDA where a central server is connected to edge devices and can orchestrate the learning process. Specifically, we assume that individual device parameters $\btheta_k$ at the lower hierarchical level are generated from a set of shared parameters at the higher hierarchical level. Through collaboratively learning these shared parameters in a federated manner, devices induce an update on their personalized parameters $\btheta_k$ that uses information from all other devices. 

Two hierarchical structures are proposed. The first defines a joint prior over $\bm{\Theta}$ parameterized by a cross-covariance matrix $\bm{\Omega}$. This allows learning a graph that achieves inductive transfer. Whereas, the second HM structure assumes that the $\btheta_k$'s are sampled from a common distribution (e.g., $\btheta_k|\bm{\phi}\sim\mathcal{N}(\bm{\mu},\bm{\Sigma})$). This allows a Bayesian treatment capable of uncertainty quantification as well as learning a global random variable $\bm{\phi}$ that can be used to predict on new unseen devices. 

We will detail our model formulations, inferences, and applications in the following two sections.



\section{A Shared Representation via Correlation}
\label{sec:HM1}



\begin{figure*}[!htbp]
    \centering
    \centerline{\includegraphics[width=0.6\columnwidth]{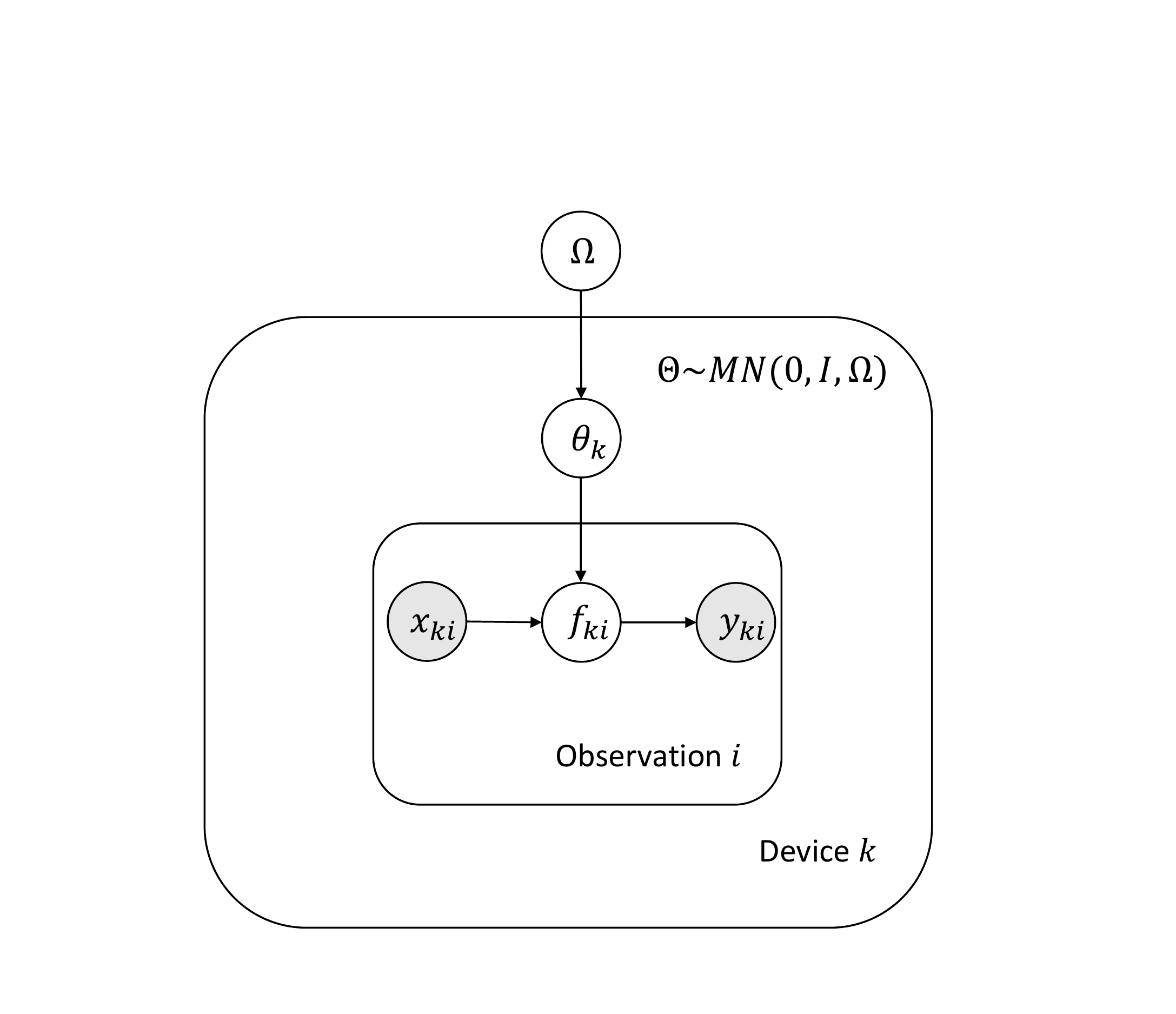}}
    \caption{The first hierarchical model structure.}
    \label{fig:hierarchical_2}
\end{figure*}

In this section, we present our first hierarchical structure (denoted as \texttt{HM1}) that establishes a shared representation by defining a structural prior over $\bm{\Theta}$ (Figure \ref{fig:hierarchical_2}). This prior is parameterized by $\bm{\Omega}$ - a $K\times K$ cross-covariance matrix. In Figure \ref{fig:hierarchical_2}, the matrix $\bm{\Omega}$ acts as a graph on the central server that encodes a shared representation among the $K$ devices and facilitates information sharing. By learning and exploiting the matrix $\bm{\Omega}$, devices can borrow information from each other to improve prediction performance. 

Mathematically,
we impose a structural prior $\mathcal{N}(\bm{0},\bm{\Omega}\otimes\bm{I})$ on $\text{Vec}(\bm{\Theta})$, where $\text{Vec}(\cdot)$ is a vectorization operation and $\otimes$ is a Kronecker product. This prior encodes the belief on the underlying distribution that generates components of $\bm{\Theta}$. More specifically, $\bm{\Omega}$ is a symmetric matrix whose $(i,j)$-th component captures the covariance between device $i$ and $j$. Overall, the aforementioned description translates to the following formulation:
\begin{align}
\label{eq:Theta}
    \bm{\Theta}\sim\mathcal{MN}(\bm{0},\bm{I},\bm{\Omega}),
\end{align}
where $\mathcal{MN}(\bm{M},\bm{A},\bm{B})$ denotes a matrix normal distribution with location (mean) parameter $\bm{M}$, row covariance $\bm{A}$ and column covariance $\bm{B}$. By Bayes' rule and incorporating Eqs. \ref{eq:y}-\ref{eq:Theta}, we can obtain the posterior distribution of $\bm{\Theta}$ as a product of the prior and the likelihood function. By omitting the constant terms and taking negative logarithm, we can obtain the negative log-likelihood function:
\begin{align*}
    -\log p(\bm{\Theta}|\bm{\Omega},\{\bm{Y}_k\}_{k=1}^K)&\propto -\log p(\{\bm{Y}_k\}_{k=1}^K|\bm{\Theta})p(\bm{\Theta}|\bm{\Omega})\\
    &=-\log \left(\prod_{k=1}^K p(\bm{Y}_k|\btheta_k)\right)-\log p(\bm{\Theta}|\bm{\Omega})\\
    &\propto\sum_{k=1}^K\frac{1}{\sigma_k^2}\norm{\bm{Y}_k-\bm{X}_k^\intercal\btheta_k}_2^2 + \text{Tr}(\bm{\Theta}\bm{\Omega}^{-1}\bm{\Theta}^\intercal)+d\log|\bm{\Omega}|\coloneqq L(\bm{\Theta},\bm{\Omega}).
\end{align*}
Therefore, our goal is to find the maximum a posteriori (MAP) of $(\bm{\Theta},\bm{\Omega})$ that minimizes the negative log-likelihood function, in a federated fashion: 
\begin{align*}
    (\bm{\Theta}^*,\bm{\Omega}^*) = \argmin_{\bm{\Theta},\bm{\Omega}} L(\bm{\Theta},\bm{\Omega}).
\end{align*}


To solve this, notice that the derivative with respect to $\bm{\Theta}$ is
\begin{align*}
    \frac{\partial L(\bm{\Theta},\bm{\Omega})}{\partial\bm{\Theta}}&=\sum_{k=1}^K\frac{-2}{\sigma_k^2}\bm{X}_k^\intercal(\bm{Y}_k-\bm{X}_k^\intercal\btheta_k) + 2\bm{\Theta}\bm{\Omega}^{-1}.
\end{align*}

In IoFT, the central server does not have access to datasets $\bm{D}=\{\bm{D}_1, \cdots, \bm{D}_K\}$, nor do devices have access to each others datasets. Further, the central server cannot share $\bm{\Omega}$ and $\bm{\Theta}$ with any device, due to the privacy constraint. Therefore, directly running gradient descent using $ \frac{\partial L}{\partial\bm{\Theta}}$ is not feasible. Yet, by scrutinizing  $\frac{\partial L}{\partial\bm{\Theta}}$, one can observe that a gradient update on each $\btheta_k$ is split into two gradients. The first term is an update from local data $\bm{D}_k$ while the second is a regularization term from all devices based on $\bm{\Omega}$. Therefore, the local parameter update can be done via the two local steps below  

\begin{align*}
    \text{Stage 1: Multiple local GD or SGD steps }& \btheta_k\leftarrow\btheta_k+2\frac{\eta_1}{\sigma_k^2}\bm{X}_k(\bm{Y}_k-\bm{X}_k^\intercal\btheta_k).\\
    \text{Stage 2: Prior Shrinkage }& \btheta_k \leftarrow \btheta_k-2\eta_2\sum_{i=1}^K\btheta_i\bm{\Omega}_{i,k}^{-1}.
\end{align*}
At the first stage, device $k$ runs multiple steps of stochastic gradient descent (SGD) or gradient descent (GD) using the local gradient information $\frac{-2}{\sigma_k^2}\bm{X}_k(\bm{Y}_k-\bm{X}_k^\intercal\btheta_k)$. To compute this gradient value, one needs to estimate the local variance parameter $\sigma_k^2$. Yet, recall that our main goal is to estimate $\btheta_k$ by borrowing information from the covariance matrix $\bm{\Omega}$. Adding to that, $\btheta_k$ and $\sigma_k^2$ are independent. Therefore, it is not necessary to estimate $\sigma_k^2$ at each local step. Here observe that $\eta_1$ is a tunable learning rate parameter and we can thus view $\eta_2\coloneqq\frac{\eta_1}{\sigma_k^2}$ as a tuning parameter in stage 1. In other words, we define $\eta_2$ as the tunable learning rate during the optimization procedure. This circumvents the need to estimate $\sigma_k^2$ locally. Nevertheless, $\sigma_k^2$ can be easily estimated from the linear residual term if it is of practitioner's interest. 

At the second stage, device $k$ will then use the aggregated information from all devices  $\sum_{i=1}^K\btheta_i\bm{\Omega}_{i,k}^{-1}$ broadcasted from the central server to update $\btheta_k$ by exploiting the covariance matrix $\bm{\Omega}$. One key notable feature of this updating framework is that the central server only needs to share an aggregated metric $\sum_{i=1}^K\btheta_i\bm{\Omega}_{i,k}^{-1}$. This operation is indeed reminiscent of federated averaging  and can preserve privacy while allowing devices to borrow strength from each other.  

Finally, we will discuss the updating rule of $\bm{\Omega}$ on the central server. The most straightforward way is to take the derivative of $L(\bm{\Theta},\bm{\Omega})$ with respect to $\bm{\Omega}$ and run gradient descent. Unfortunately, $L(\bm{\Theta},\bm{\Omega})$ is not convex with respect to $\bm{\Omega}$ due to $d\log|\bm{\Omega}|$, despite being convex in terms of $\text{Tr}(\bm{\Theta}\bm{\Omega}^{-1}\bm{\Theta}^\intercal)$. However, seminal work from \citet{zhang2012convex} suggests that when minimizing $\text{Tr}(\bm{\Theta}\bm{\Omega}^{-1}\bm{\Theta}^\intercal) + d\log|\bm{\Omega}|$ with respect to $\bm{\Omega}$, the term $d\log|\bm{\Omega}|$ can be simply replaced by a simple convex constraint of $\text{Tr}|\bm{\Omega}|=1$. By doing so, a closed-form solution is directly derived $\bm{\Omega}\leftarrow \frac{(\bm{\Theta}^\intercal\bm{\Theta})^{0.5}}{\text{Tr}((\bm{\Theta}^\intercal\bm{\Theta})^{0.5})}$. Here one can view that $\bm{\Theta}^\intercal\bm{\Theta}$ encodes the information of device covariance. Unfortunately, this approach typically faces singularity issues when $\bm{\Theta}^\intercal\bm{\Theta}$ is not a positive definite matrix (e.g., contains zero elements). To resolve this, we propose a new updating procedure that prevents an abrupt change in $\bm{\Omega}$ to safeguard against singularity. More specifically, we have
\begin{align*}
    \bm{\Omega}\leftarrow (1-\alpha)\bm{\Omega} + \frac{\alpha}{d}\bm{\Theta}^\intercal\bm{\Theta}.
\end{align*}
Here, $\alpha$ is a parameter that controls the change in $\bm{\Omega}$ and $\frac{\bm{\Theta}^\intercal\bm{\Theta}}{d}$ encodes the devices' covariance. A small $\alpha$ renders a conservative updating rule while a large $\alpha$ under-weights the importance of the covariance matrix from the previous communication round.

Here note that, in the aforementioned framework, the central server selects all devices at each communication round. This scheme is known as full device participation. In reality, however, some local devices are often offline or unwilling to respond due to various reasons. To accommodate this situation, one can sample a subset of devices at each communication round. We term this scenario as partial device participation. We summarize the detailed algorithm in Algorithm \ref{algo:cor}. 



\begin{algorithm}[!htbp]
	\SetAlgoLined
	\KwData{Number of devices $K$, Set $\mathcal{S}$ that contains indices of the selected devices, number of communication rounds $C$, randomly initialized model parameter $\bm{\Theta}$, initial matrix $\bm{\Omega}=\bm{I}$, learning rate $\eta_2$ (selected by grid-search or other tuning methods), proportion $\alpha=0.1$ (default), dimension $d$.}
	\For{$c=0:(C-1)$}{
	    Server broadcasts column of $\bm{\Theta}$ (i.e., $\btheta_k$)\;
	    \For{$k\in\mathcal{S}$}{
	        \For{$t=0:(T-1)$}{
	            Device-side: (Sampling Batch) Sample a subset of data $(\bm{X}^b_k,\bm{Y}^b_k)$ from $\bm{D}_k$, where superscript $b$ means batch\;
	            Device-side: (SGD or GD) $\btheta_k^{(t+1)}=\btheta_k^{(t)}+2\eta_2\bm{X}_k^{b\intercal}(\bm{Y}^b_k-\bm{X}_k^{b\intercal}\btheta_k^{(t)})$\;
	        }
	        Device-side: (Prior Shrinkage) $\btheta_k \leftarrow \btheta_k-2\eta_2\sum_{i=1}^K\btheta_i\bm{\Omega}_{i,k}^{-1}$\;
		}
		Server-side: $\bm{\Omega}\leftarrow (1-\alpha)\bm{\Omega} + \frac{\alpha}{d}\bm{\Theta}^\intercal\bm{\Theta}$.
	}
	Return $\bm{\Theta},\bm{\Omega}$.
	\caption{Improving Device Performance by Exploiting Structural Covariance}
	\label{algo:cor}
\end{algorithm}

As we will show in our numerical studies, this simple to implement algorithm requires very few communication rounds to recover the true parameters and excels at leveraging knowledge across all devices.

\section{A Hierarchical Model based on the Distribution Assumption}
\label{sec:HM2}

So far, we have presented \texttt{HM1} which exploits the relationship among devices to improve prediction performance. One drawback of Algorithm \ref{algo:cor} is that it only returns a point estimate of $\bm{\Theta}$. In practice, it is also desirable to quantify the uncertainty in the parameter estimates. Additionally, the estimated $\bm{\Theta}$ and $\bm{\Omega}$ cannot provide any borrowable information for new devices, yet the idea of fast adaptation to new unseen data is crucial in many fields such as meta-learning \citep{vanschoren2019meta}. In this section, we will present an alternative model structure that is formulated from a Bayesian perspective to tackle the aforementioned issues. 

\subsection{Structure and Formulation}
\label{subsec:for}

\begin{figure*}[!htbp]
    \centering
    \centerline{\includegraphics[width=0.6\columnwidth]{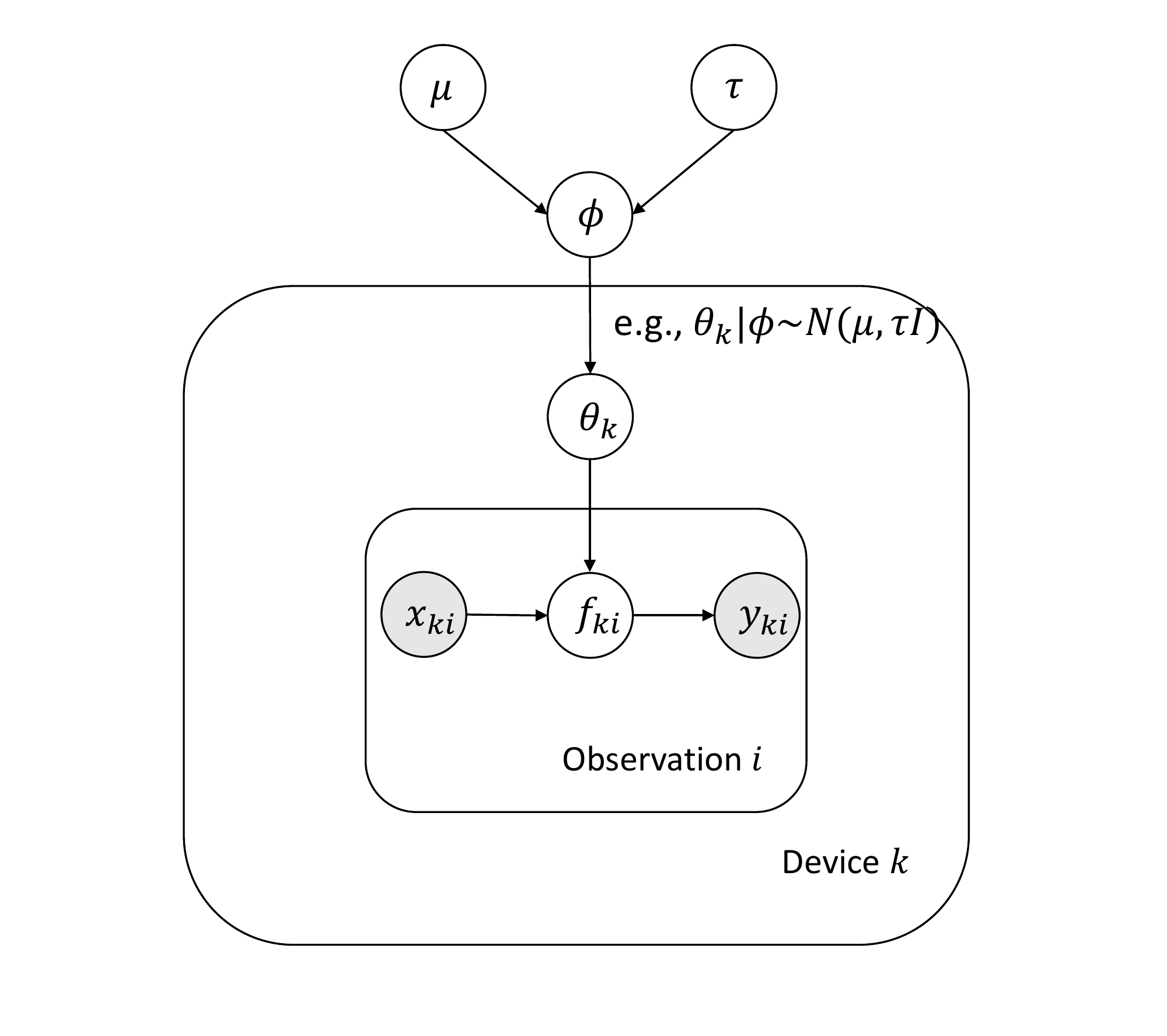}}
    \caption{The second hierarchical model structure.}
    \label{fig:hierarchical_1}
\end{figure*}

Our second structure (denoted as \texttt{HM2}) assumes all device parameters are generated from the same underlying distribution. To give a simple example, one can assume $\btheta_k|\bm{\phi}\sim\mathcal{N}(\bm{\mu},\tau\bm{I})$ (Figure \ref{fig:hierarchical_1}) where $\bm{\phi}=(\bm{\mu},\tau)$ is a set of a global hyper-parameters on the central server. Here it is critical to note that $\tau\bm{I} \in \mathbb{R}^{d \times d}$ is a within covariance matrix and does not denote covariances across $K$ devices as all $\theta_k$'s come from the same underlying distribution.  This assignment indicates that $\{\btheta_k\}_{k=1}^K$ are related and generated from the same distribution, yet the degree of model similarity is controlled by the variance parameter $\tau$. A small $|\tau|$ implies all model parameters are similar (i.e., homogeneous) and the hierarchical model is closely related to learning a single common parameter $\btheta$ that fits all devices' data. On the other hand, a large variance $|\tau|$ incurs more heterogeneity among devices. In the extreme case when $|\tau|\to\infty$, the hierarchical model is equivalent to a separate modeling approach where each device's data is fitted separately \citep{albert2019probability}.

Now, we formally define the \texttt{HM2} formulation. From our hierarchical definition and taking a fully Bayesian treatment by placing a prior $p(\bm{\phi})$ on the global hyper-parameters, the joint posterior of $\{\btheta_k\}_{k=1}^K$ and $\bm{\phi}$ for \texttt{HM2} can be written as:
\begin{align*}
    p(\btheta_1,\ldots,\btheta_K,\bm{\phi}|\bm{Y}_1,\ldots,\bm{Y}_K)&\propto p(\bm{Y}_1,\ldots,\bm{Y}_K|\btheta_1,\ldots,\btheta_K,\bm{\phi})p(\btheta_1,\ldots,\btheta_K,\bm{\phi})\\
    &=p(\bm{\phi})\prod_{k=1}^Kp(\bm{Y}_k|\btheta_k,\bm{\phi})p(\btheta_k|\bm{\phi}) \, .
\end{align*}
To further contextualize \texttt{HM2} we provide an example of a possible formulation.


\begin{exmp}
Assume each device $k$ fits a linear regression parameterized by $\btheta_k$. The local dataset is given as $(\bm{X}_k,\bm{Y}_k)$ for all $k$. Then one possible hierarchical formulation is:
\begin{align*}
    \bm{Y}_k|\btheta_k,\bm{\phi}&\sim\mathcal{N}(\bm{X}_k^\intercal\btheta_k,\sigma_k^2\bm{I})\\
    \btheta_k|\bm{\phi}&\sim\mathcal{N}(\bm{\mu},\text{diag}(\tau_1,\ldots,\tau_d))\\
    \bm{\mu}&\sim\mathcal{N}(\bm{0},\bm{I})\\
    \tau_i&\sim\log\mathcal{N}(0,1), \forall i\\
    \bm{\phi} &= (\bm{\mu},\tau_1,\ldots,\tau_d),
\end{align*}
where $\text{diag}(\tau_1,\ldots,\tau_d)$ defines a diagonal matrix with components $\tau_1,\ldots,\tau_d$.
\end{exmp}

Clearly, inferring the joint posterior above is very challenging in a federated setting. Yet, in a hierarchical model, if we know the posterior over the upper hierarchical level $p(\bm{\phi}|\{\bm{Y}_k\}_{k=1}^K)$ (i.e. over global hyper-parameters $\bm{\phi}$), we can directly use this posterior as a prior to infer the lower level $p(\btheta_k|\{\bm{Y}_k\}_{k=1}^K)$ parameters locally. 

More specifically, we can derive that
\begin{align*}
    p(\bm{\phi}|\bm{Y}_1,\ldots,\bm{Y}_K)=p(\bm{\phi})\prod_{k=1}^K\int p(\bm{Y}_k|\btheta_k,\bm{\phi})p(\btheta_k|\bm{\phi})d\btheta_k=p(\bm{\phi})\prod_{k=1}^Kf_k(\bm{\phi}),
\end{align*}
where $f_k(\bm{\phi})=\int p(\bm{Y}_k|\btheta_k,\bm{\phi})p(\btheta_k|\bm{\phi})d\btheta_k$. As a consequence, our main goal is to collaboratively learn $p(\bm{\phi}|\bm{Y}_1,\ldots,\bm{Y}_K)$ in a federated setting. One key challenge, however, is that the central server does not have any access to any edge dataset $\bm{D}_k$ and therefore directly computing $p(\bm{\phi}|\bm{Y}_1,\ldots,\bm{Y}_K)$ is infeasible. For this reason we resort to a trick based on approximate inference methods to learn this posterior distribution.

\subsection{Federated Bayesian Inference - Expectation Propagation}





In this section, we will present the federated inference framework for \texttt{HM2}. Specifically, our goal is to learn the posterior density $p(\bm{\phi}|\bm{Y}_1,\ldots,\bm{Y}_K)$. In statistics, the most straightforward and popular approaches to do so are Markov chain Monte Carlo (MCMC) methods. Yet, as will be clear shortly, we argue that sampling methods are not practical in the federated hierarchical setting due to their sequential nature. Take Gibbs sampling as an example, device 1 needs to sample $\btheta_1$ from the density $p(\btheta_1|\btheta_2,\ldots,\btheta_K,\bm{\phi},\bm{Y}_1)$ then passes those samples to the central server. The central sever then needs to transmit those sampled $\btheta_1$ to device 2, and device 2 will sample from $p(\btheta_2|\btheta_1,\ldots,\btheta_K,\bm{\phi},\bm{Y}_2)$. It can be seen that this sequential nature of MCMC significantly increases the communication cost and also slows down the federated optimization process when the total number of device is large. Even if one can smartly parallelize the sampling process, the number of MCMC samples obtained locally will be large if the dimension is high due to the curse of dimensionality \citep{jordan2018communication}.

To resolve the aforementioned issues, we resort to expectation propagation (EP) \citep{minka2001family} to approximate the posterior distribution. EP is one of the most widely-used algorithms for computing an approximate posterior distribution \citep{minka2013expectation, vehtari2020expectation}. Here, we first briefly introduce the idea of EP in a centralized regime. Consider a posterior distribution with independent data points
\begin{align*}
    \pi(\bm{\phi})\coloneqq p(\bm{\phi}|\bm{Y})\propto p(\bm{\phi})\prod_{i=1}^N p(y_i|\bm{\phi})
\end{align*}
where $\bm{Y}=(y_1,\ldots,y_N)^\intercal$ is the data vector. EP approximates $\pi(\bm{\phi})$ by a density $q(\bm{\phi})$ such that
\begin{align*}
    q(\bm{\phi}) = p(\bm{\phi})\prod_{i=1}^N q_i(\bm{\phi}).
\end{align*}
Intuitively, EP uses $q_i(\bm{\phi})$ to approximate $p(y_i|\bm{\phi})$, for all $i$. To achieve this goal, at each iteration, EP first takes an approximation factor $q_i(\bm{\phi})$ out from the current $q(\bm{\phi})$ and replaces it by the true factor $p(y_i|\bm{\phi})$. This step yields a new density $q^{new}(\bm{\phi})$. This resulting new density can be used as an updated approximated posterior. This step is iterated over all $i$ till convergence. Please refer to \cite{youtube2016} for a comprehensive summary of EP. It can be seen that EP can be naturally extended to FDA where each device can be viewed as an independent ``data point". In the following paragraphs, we will detail the federated extension of EP.

The main idea is to approximate terms $f_k(\bm{\phi})$ by a local device approximation function $q_k(\bm{\phi})$ for all $k=1,\ldots,K$. More specifically, we have
\begin{align*}
    p(\bm{\phi}|\bm{Y}_1,\ldots,\bm{Y}_K)\approx p(\bm{\phi})\prod_{k=1}^K q_k(\bm{\phi})\coloneqq q(\bm{\phi}).
\end{align*}
Using the framework of EP, we gradually update $q(\bm{\phi})$ by iteratively renewing $q_k(\bm{\phi})$ at each device $k$. During each communication round, given the estimated $q(\bm{\phi})$ broadcasted from the server, device $k$ first computes the cavity distribution
\begin{align}
\label{eq:cav}
    q_{-k}(\bm{\phi})\propto\frac{q(\bm{\phi})}{q_{k}(\bm{\phi})}
\end{align}
and the tilted distribution
\begin{align}
\label{eq:til}
    q_{\backslash k}(\bm{\phi})\propto f_k(\bm{\phi})q_{-k}(\bm{\phi}).
\end{align}
It then computes the updated posterior approximation such that
\begin{align}
\label{eq:new_eq}
    q^{new}(\bm{\phi})= q_{\backslash k}(\bm{\phi}).
\end{align}
Intuitively, the cavity distribution $q_{-k}(\bm{\phi})$ removes the impact of the old $q_k(\bm{\phi})$ from the approximated posterior density $q(\bm{\phi})$ and the tilted distribution adds the true target density $f_k(\bm{\phi})$ to $q_{-k}(\bm{\phi})$. As a result, we use the tilted distribution as an updated approximation to the posterior density of $\bm{\phi}$.



Afterwards, device $k$ calculates the change in its local approximation by
\begin{align}
\label{eq:change}
    \Delta q_k(\bm{\phi})=\frac{q^{new}(\bm{\phi})}{q(\bm{\phi})}.
\end{align}
Instead of sending $q^{new}$ back to the server, we calculate the change in the global posterior imposed by device $k$ via Eq. \eqref{eq:change} and sends $\Delta q_k(\bm{\phi})$ to the central server. The server aggregates all device approximations by
\begin{align}
\label{eq:agg}
    q(\bm{\phi})\leftarrow q(\bm{\phi})\prod_{k=1}^K\Delta q_k(\bm{\phi}).
\end{align}

We summarize the EP algorithm in Algorithm \ref{algo:Fed-EP}.

\begin{algorithm}[!htbp]
	\SetAlgoLined
	\KwData{number of devices $K$, Set $\mathcal{S}$ that contains indices of the selected devices, number of communication rounds $C$, initial approximation $\{q_k(\bm{\phi})\}_{k=1}^K$, prior $p(\bm{\phi})$, learning rate $\eta$ (Selected by grid-search)}
	\For{$c=0:(C-1)$}{
	    Server broadcasts $q(\bm{\phi})$\;
	    \For{$k\in\mathcal{S}$}{
	        Device-side: Calculate the cavity distribution $q_{-k}(\bm{\phi})$ using Eq. \eqref{eq:cav}\;
	        Device-side: Calculate the tilted distribution $q_{\backslash k}(\bm{\phi})$ using Eq. \eqref{eq:til}\;
	        Device-side: Get new $q(\bm{\phi})$ from the tilted distribution using Eq. \eqref{eq:new_eq}\;
	        Device-side: Calculate $\Delta q_k(\bm{\phi})$ using Eq. \eqref{eq:change} and update local $q_k(\bm{\phi})$ \;
	        Device-side: Send $\Delta q_k(\bm{\phi})$ to the central server\;
		}
		Server-side: Update $q(\bm{\phi})$ using Eq. \eqref{eq:agg}\;
	}
	Return $q(\bm{\phi})$.
	\caption{The Federated Expectation Propagation Algorithm}
	\label{algo:Fed-EP}
\end{algorithm}

\subsubsection{Posterior of Device Parameters} 

Once we obtain $q(\bm{\phi})$ that approximates $p(\bm{\phi}|\bm{Y}_1,\ldots,\bm{Y}_k)$, we can further estimate the posterior of device parameters. Specifically, given a device $k$,
\begin{align*}
    p(\btheta_k|\{\bm{Y}_k\}_{k=1}^K)&=\int_{\bm{\phi}}\int_{\btheta_j,j\neq k} p(\btheta_k,\bm{\phi}|\{\bm{Y}_k\}_{k=1}^K) d\btheta_jd\bm{\phi}\\
    &\propto\int p(\bm{\phi})p(\bm{Y}_k|\btheta_k,\bm{\phi})p(\btheta_k|\bm{\phi})\prod_{j\neq k}\int p(\bm{Y}_j|\btheta_j,\bm{\phi})p(\btheta_j|\bm{\phi})d\btheta_jd\bm{\phi}\\
    &\approx \int q_{-k}(\bm{\phi})p(\bm{Y}_k|\btheta_k,\bm{\phi})p(\btheta_k|\bm{\phi})d\bm{\phi}.
\end{align*}
As a consequence, we can use the posterior of $\btheta_k$ to quantify uncertainties or conduct hypothesis testing. The posterior samples from $p(\btheta_k|\{\bm{Y}_k\}_{k=1}^K)$ can be obtained by off-the-shelf posterior sampling techniques (see \texttt{mcmc} package in R or \texttt{NumPyro} library in Python). Here we provide a simpler sampling trick. In the above equation, if we ignore the integral, we can obtain the joint posterior 
\begin{align}
\label{eq:posterior}
    p(\btheta_k,\bm{\phi}|\{\bm{Y}_k\}_{k=1}^K)\approx  q_{-k}(\bm{\phi})p(\bm{Y}_k|\btheta_k,\bm{\phi})p(\btheta_k|\bm{\phi}).
\end{align}
As a result, one can ignore the integration and use sampling methods to jointly sample ($\btheta_k,\bm{\phi}$) from Eq. \eqref{eq:posterior} and discard $\bm{\phi}$. Now, given $M$ samples $\{\btheta_{ki}\}_{i=1}^M$, we can readily use the samples to estimates moments, coverage probability and do hypothesis tests. 

Here we should note that the estimated posterior density $q(\bm{\phi})$ encodes crucial information across all devices (recall the explanation in Sec. \ref{subsec:for}). One can exploit this information to for a new device to achieve fast adaption. For example, we can treat the posterior mean of $\bm{\phi}$ as an initial model parameter for a new device. This idea is similar to meta-learning \citep{vanschoren2019meta}, where one tries to learn a global model that can quickly adapt to a new task.

\subsubsection{Normal Approximation} 

In practice, it is common to model $p(\bm{\phi})$ and $q_k(\bm{\phi}),\forall k$ as normal densities. This is due to a very useful property of normal random variables.
\begin{lemma}
\label{lemma:1}
\citep{williams2006gaussian} Suppose there are two normal random variables (with the same dimension) such that $\btheta_1\sim\mathcal{N}(\bm{\mu}_1,\bm{\Sigma}_1)$ and $\btheta_2\sim\mathcal{N}(\bm{\mu}_2,\bm{\Sigma}_2)$. Let $\bm{r}_i=\bm{\Sigma}_i^{-1}\bm{\mu}_i$, $\bm{Q}_i=\bm{\Sigma}_i^{-1}$ for $i=1,2$. Define $p(\btheta_+)= p(\btheta_1)p(\btheta_2)$ and $p(\btheta_-)=\frac{ p(\btheta_1)}{p(\btheta_2)}$. We have that
\begin{align*}
    \btheta_+\sim\mathcal{N}(\bm{r}_1+\bm{r}_2,\bm{Q}_1+\bm{Q}_2)\\
    \btheta_-\sim\mathcal{N}(\bm{r}_1-\bm{r}_2,\bm{Q}_1-\bm{Q}_2).
\end{align*}
\end{lemma}
Using Lemma \ref{lemma:1}, one can efficiently implement the EP algorithm. Here, we detail the implementation technique. We model the prior of $\bm{\phi}$ as a multivariate normal random variable with mean $\bm{\mu}_0$ and variance $\bm{\Sigma}_0$. We also assume $g_k(\bm{\phi})$ has a normal density parameterized by $\bm{\mu}_k,\bm{\Sigma}_k$, for all $k$. If the support of some components in $\bm{\phi}$ do not lie in $\mathbb{R}$, one can always perform a logarithmic or logistic transformation to those components. By Gaussian properties, Eq. \eqref{eq:agg} can be computed in closed-form such that $g(\bm{\phi})$ has a normal density parametrized by mean $\bm{r}_0+\sum_{k=1}^K\bm{r}_k$ and variance $\bm{Q}_0+\sum_{k=1}^K\bm{Q}_k$, where $\bm{r}_j=\bm{\Sigma}_j^{-1}\bm{\mu}_j$ and $\bm{Q}_j=\bm{\Sigma}_j^{-1}$, for all $j=0,1,\ldots,K$. Similarly, the cavity distribution in Eq. \eqref{eq:cav} can be computed in a closed-form by a subtraction operation. Compared to sampling approaches, one key advantage of EP is that the computation and communication steps are simple and efficient. The central server and devices only need to transmit the mean vector and variance matrix to perform model updating and aggregation. We detail this idea in Algorithm \ref{algo:Fed-Gaussian}.

\begin{algorithm}[!htbp]
	\SetAlgoLined
	\KwData{number of devices $K$, Set $\mathcal{S}$ that contains indices of the selected devices, number of communication rounds $C$, initial approximation $\{\bm{r}_k,\bm{Q}_k\}_{k=1}^K$, prior $\bm{r}_0,\bm{Q}_0$, initial posterior parameters $\bm{r}=\bm{r}_0+\sum_{k=1}^K\bm{r}_k,\bm{Q}=\bm{Q}_0+\sum_{k=1}^K\bm{Q}_k$, learning rate $\eta$ (Selected by grid-search)}
	\For{$c=0:(C-1)$}{
	    Server broadcasts $\bm{r},\bm{Q}$\;
	    \For{$k\in\mathcal{S}$}{
	        Device-side: Calculate the cavity distribution with parameters $\bm{r}_{-k}\coloneqq\bm{r}-\bm{r}_k,\bm{Q}_{-k}\coloneqq\bm{Q}-\bm{Q}_k$\;
	        Device-side: Calculate the tilted distribution $\bm{r}_{\backslash k},\bm{Q}_{\backslash k}$\;
	        Device-side: Obtain new $\bm{r}_k^{new},\bm{Q}_k^{new}$\;
	        Device-side: Calculate $\Delta \bm{r}_k=\bm{r}_k^{new}-\bm{r}_k,\Delta\bm{Q}_k=\bm{Q}_k^{new}-\bm{Q}_k$ \;
	        Device-side: Send $\Delta \bm{r}_k,\Delta\bm{Q}_k$ to the central server\;
		}
		Server-side: Update $\bm{r}=\bm{r}+\sum_{k\in\mathcal{S}}\Delta\bm{r}_k,\bm{Q}=\bm{Q}+\sum_{k\in\mathcal{S}}\Delta\bm{Q}_k$\;
	}
	Return $\bm{r}, \bm{Q}$.
	\caption{The Federated Expectation Propagation Algorithm using Normal Approximation}
	\label{algo:Fed-Gaussian}
\end{algorithm}

\subsection{Federated \& Penalized Regression for Variable Selection}
\label{sec:HM2-variable}

To move a step further, we ask ``is it possible to let devices exploit the shared representation structure to perform variable selection?" Indeed, there are some attempts to tackle this question from a frequentist perspective. \cite{yuan2021federated} develop a federated composite optimization framework that solves the federated lasso problem. \cite{tong2020federated} propose a federated iterative hard thresholding algorithm to tackle non-convex penalized regression. Despite these few efforts in exploring variable selection from a frequentist perspective, no literature exists in the Bayesian setting.



\citet{tibshirani1996regression} has shown that a Lasso estimate can be achieved when the regression parameters have $\textit{i.i.d.}$ Laplace priors. Since then, researchers have started to build Bayesian priors for many other penalized regressions such as the elastic net and fussed Lasso. Please refer to the work of \cite{van2019shrinkage} for a detailed literature review. Inspired by the Bayesian interpretation of penalized regressions, we develop a hierarchical structure, based upon \texttt{HM2}, to perform federated variable selection. To proceed, we impose priors on $\btheta_k$, for all $k$, such that
\begin{align*}
    \theta_{ki}|\bm{\phi}&\sim \pi(\lambda,\sigma^2), \forall i=1,\ldots,d
\end{align*}
where $\bm{\phi}=(\lambda,\sigma^2)$, $\lambda$ is a regularization parameter and $\sigma^2$ is a variance parameter. Here, $\pi(\lambda,\sigma^2)$ is a distribution parameterized by $\lambda,\sigma^2$. For instance, if we set $\pi(\lambda,\sigma^2)$ to be a Laplace distribution with zero mean and $\frac{\sigma}{\lambda}$ diversity, then we recover Lasso regression \citep{tibshirani1996regression}. Another example is if we set $\pi(\lambda,\sigma^2)$ to be $\mathcal{N}(0,\frac{\sigma}{\lambda})$, then we recover Ridge regression. There are many possible choices of prior beliefs on $\sigma^2$ and $\lambda$. In this work, we impose log-normal priors on $\sigma^2$ and $\lambda$ \citep{van2019shrinkage} and we set $\bm{\phi}=(\log\lambda,\log\sigma^2)$. Our framework can flexibly incorporate other priors such as a non-informative prior on $\sigma^2$ or a half-Cauchy prior on $\lambda$. In this work, we will use a log-normal prior as an illustrative example.

The posterior distribution of $\btheta_k$, for all $k$, and $\bm{\phi}$ can be learned by Algorithm \ref{algo:Fed-EP}. Here, one caveat is that, unlike frequentist penalized regressions, the Bayesian methods do not shrink regression coefficients to be exactly zero. As a result, we will calculate the credible interval (CI) for each parameter. If the CI of a parameter, say $\theta_{ki}$, covers $0$, we will exclude this predictor.

\section{Simulation Studies}
\label{sec:sim}


\subsection{HM1 Proof of Concept using Algorithm \ref{algo:cor}}

\textbf{Case I:} We assume that $K=2$ and generate $(\btheta_1, \btheta_2)$ from a matrix-variate normal distribution with zero mean and $\bm{I}\otimes\bm{\Omega}=\bm{I}\otimes\begin{bmatrix}1&0.7\\0.7&1\end{bmatrix}$ covariance. The input space dimension is $d=5$. Data on each device are generated from linear models using the generated parameters. We set noise to be $0.05$. To demonstrate the benefits of our correlation based construction in HM1, we create imbalanced sample sizes on the devices. Specifically, device 1 only has $N_1=20$ data points and device 2 has $N_2=200$ data points. We train Algorithm \ref{algo:cor} with $C=30, \eta_2=0.01,\alpha=0.1$ and we set the number of local steps $T$ to be $20$. We compare the performance of Algorithm \ref{algo:cor} with a separate modeling approach where each device fits its own model without communication. Specifically, each device runs $600$ local SGD steps with learning rate $0.01$.

\textbf{Case II:} We set $K=100$ and generate a $100\times100$ positive definite matrix $\bm{\Omega}$ using the R package \texttt{clusterGeneration}. We then generate true device parameters based on the matrix $\bm{\Omega}$. We set $d=8$ and generate data using the linear models with noise $0.1$. For the first $30$ devices, we assign $40$ data points and for the remaining $70$ devices, we assign $275$ data points. Overall, we create an imbalanced data generation scenario. Case II can be viewed as a generalization of Case I with more devices. Again, we train models using the same hyperparameters as those in Case I.

\textbf{Case III:} We use the same setting as the one in Case II but all devices have $20$ data points (i.e., balanced data generation).

\textbf{Case IV:} We increase the sample size on each device to $200$ and use the same setting as the one in Case III.

\textbf{Performance Evaluation:} Denote by $(\bm{X}^{*}_{k},\bm{Y}^{*}_{k})$ the testing dataset on device $k$ where $\bm{X}^{*}_{k}=[x^{*}_{k1},\ldots,x^{*}_{kN_k^*}]^\intercal$ and $\bm{Y}^{*}_{k}=[y^*_{k1},\ldots,y^*_{kN^*_k}]$. The averaged RMSE across all devices is defined as
\begin{align*}
    \text{A-RMSE} = \frac{1}{K}\sum_{k=1}^K\sqrt{\frac{\sum_{i=1}^{N^*_k}(f(x^{*}_{ki})-y^*_{ki})^2}{N^*_k}}.
\end{align*}
On each device, we generate $1,000$ data points using the true device parameters for testing. In Case I, the RMSE is calculated on device $1$. In Case II, the A-RMSE is calculated using devices $k\in\{1,\ldots,30\}$. Here our goal is to assess the prediction accuracy on devices with scant data. In Cases III/IV, we calculate A-RMSE using all $100$ devices. We report our results in Table \ref{table:case123}. It can be seen that Algorithm \ref{algo:cor} yields much smaller A-RMSE under the imbalanced data scenario. This conveys the importance of borrowing strength from other devices under the FDA framework. In Case III, the local sample size is not enough such that each model alone cannot perform well. However, our FDA can still benefit devices' training by borrowing information from other devices. In case IV, Algorithm \ref{algo:cor} does not offer a major improvement as all local devices have enough data. In this case, doing local training without collaboration should be sufficient.

\begin{table}[!htbp]
\centering
\begin{tabular}{ccc}
\hline
 Case & Algorithm \ref{algo:cor} & Separate   \\ \hline
 I &  $0.081 (\pm 0.001)$ & $0.094 (\pm 0.001)$ \\ \hline
 II & $0.050 (\pm 0.000)$ & $0.056 (\pm 0.001)$ \\ \hline
 III & $0.044 (\pm 0.002)$ & $0.072 (\pm 0.004)$ \\ \hline
 IV & $0.035 (\pm 0.000)$ & $0.035 (\pm 0.000)$ \\ \hline
\end{tabular}
\caption{The A-RMSE of our proposed model and the separate model over 30 independent runs. We report standard deviation in brackets.}
\label{table:case123}
\end{table}

\begin{figure*}[!htbp]
    \centering
    \centerline{\includegraphics[width=0.8\columnwidth]{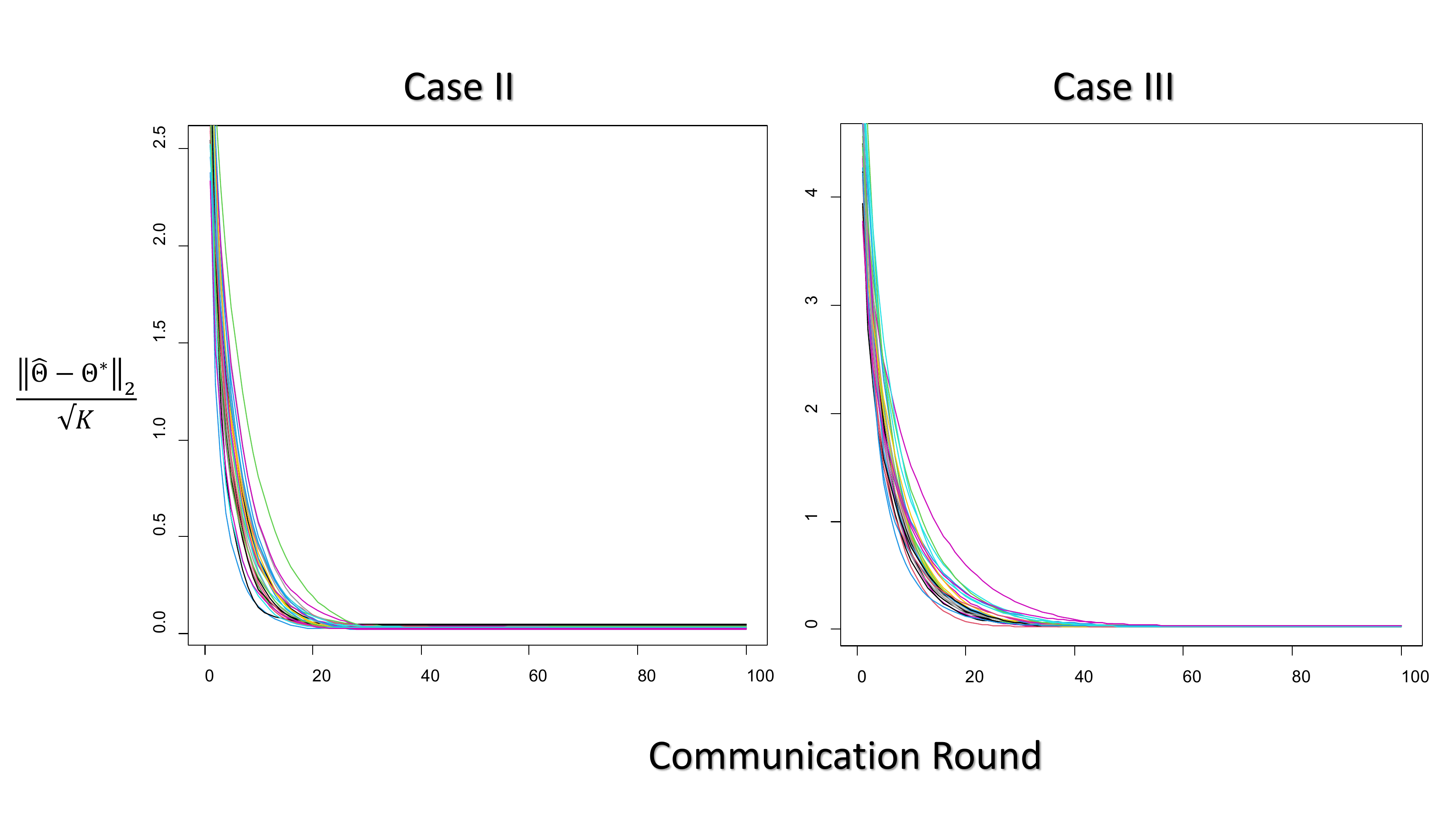}}
    \caption{Plot of $\frac{\norm{\hat{\bm{\Theta}}-\bm{\Theta}^*}}{\sqrt{K}}$ versus communication round. Each color represents one independent run.}
    \label{fig:case_23}
\end{figure*}

\textbf{Accuracy of Parameter Estimation:} Denote by $\hat{\bm{\Theta}}$ the concatenated estimated device parameters and $\bm{\Theta}^*$ the true data-generating parameters. In Figure \ref{fig:case_23}, we plot $\frac{\norm{\hat{\bm{\Theta}}-\bm{\Theta}^*}}{\sqrt{K}}$ versus communication round for Case II and III over 30 independent runs. It can be seen that Algorithm \ref{algo:cor} accurately recovers the true underlying model parameters. Furthermore, it can be observed that Algorithm \ref{algo:cor} typically converges within 30-40 communication rounds. We observed that, in all simulations, Algorithm \ref{algo:cor} can be trained within 5 seconds on a standard laptop. In conclusion, our proposed algorithm is easy to implement and optimize.

\subsection{HM2 Proof of Concept}

In this section, we test the variable selection performance of \texttt{HM2}. Following the examples in \cite{van2019shrinkage}, we create several simulation cases below:

\textbf{Case I:} We set $K=10$, $\btheta_{true}=(3, 1.5, 0, 0, 2, 0, 0, 0)^\intercal$ and generate all columns of $\{\bm{X}_k\}_{k=1}^K$ from a standard multivariate normal distribution. We then generate $\{\bm{Y}_k\}_{k=1}^K$ using $\btheta_{true}$ and $\{\bm{X}_k\}_{k=1}^K$ for all $k$. We set noise to be 0.05. Each device has $100$ data points for training and $1000$ data points for testing.

\textbf{Case II:} We use the same setting as the one in Case I. The difference is that the first 2 devices only have $20$ data points each while the other devices have $200$ data points each. The number of testing data points is $1000$.

\textbf{Case III:} We set $K=20$, $\btheta_{true}=(\underbrace{3,\ldots,3,}_{10}\underbrace{0,\ldots,0,}_{10}\underbrace{3,\ldots,3}_{10})^\intercal$. Each device has $40$ observations for training and $400$ observations for testing.

We evaluate the performance of our model based on prediction and variable selection accuracy. The prediction accuracy is evaluated by A-RMSE. Variable selection accuracy is based on the averaged correct and false inclusion rates. To decide whether to include a variable or not, we first calculate a $90\%$ credible interval for each parameter. If the CI covers $0$, we will exclude this predictor. Results are reported in Table \ref{table:sim_variable}. It can be seen that our proposed federated variable selection methods can correctly identify more than $85\%$ of effective predictors while maintaining low false inclusion rates.

\begin{table}[!htbp]
\centering
\small
\begin{tabular}{cccc}
\hline
Methods & A-RMSE & Averaged Correct Inclusion Rate & Averaged False Inclusion Rate\\ \hline
Lasso (\texttt{HM2}, Case I) & $0.055 (\pm0.001)$ & $0.880 (\pm0.003)$ & $0.095 (\pm0.001)$\\ \hline
Lasso (\texttt{HM2}, Case II) & $0.062 (0.002)$ & $0.875 (\pm0.004)$ & $0.101 (\pm0.001)$\\ \hline
Lasso (\texttt{HM2}, Case III) & $0.088 (0.001)$ & $0.891 (\pm0.005)$ & $0.115 (\pm0.001)$\\ \hline
\end{tabular}
\caption{The A-RMSE and averaged correct/false inclusion rates for different federated variable selection methods, over 30 experimental runs.}
\label{table:sim_variable}
\end{table}

As mentioned in Sec. \ref{sec:HM2}, one advantage of \texttt{HM2} is that it can provide uncertainty quantification (UQ) for parameter estimation. We will provide two examples to demonstrate the UQ capability of \texttt{HM2}. 

\begin{figure*}[!htbp]
    \centering
    \centerline{\includegraphics[width=\columnwidth]{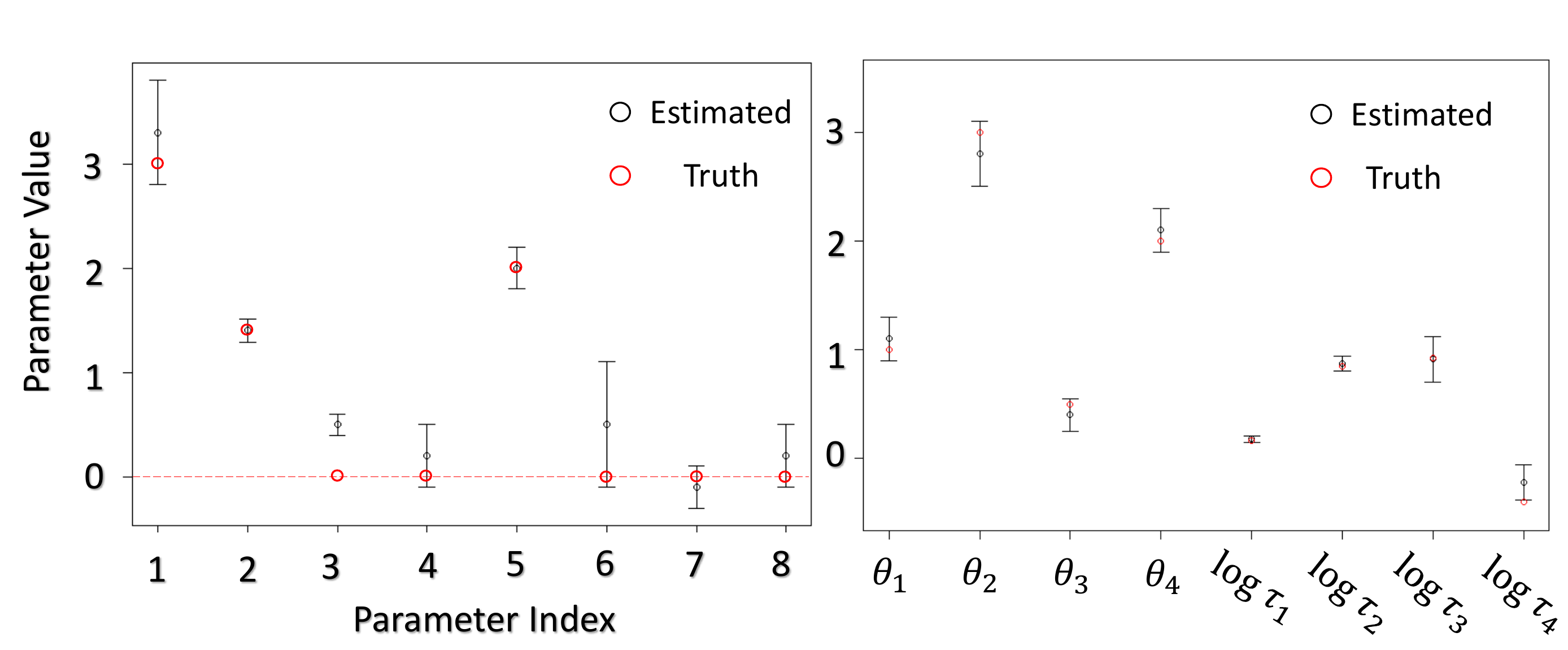}}
    \caption{Plot of parameter estimation and confidence interval.}
    \label{fig:CI}
\end{figure*}

1. We collect the estimated posterior for parameters $\btheta_1$ from an independent run in case I and calculate the mean and $90\%$ credible interval. The resulting plot is presented in Figure \ref{fig:CI} (Left). It can be seen that the true parameters are included in the confidence interval generated by \texttt{HM2}.

2. We create $K=100$ devices and generate device parameter
    \begin{align*}
        \btheta_k|\bm{\phi}\sim\mathcal{N}(\bm{\mu}_{true},\bm{\Sigma}_{true})\coloneqq\mathcal{N}\left(
        \begin{bmatrix} 1\\ 3 \\ 0.5 \\ 2\end{bmatrix},
        \begin{bmatrix} 
        1.17 & 0 & 0 & 0 \\ 
        0 & 2.35 & 0 & 0 \\
        0 & 0 & 2.52 & 0 \\
        0 & 0 & 0 & 0.67
        \end{bmatrix}
        \right)
    \end{align*}
    for $k\in\{1,\ldots,100\}$. We then use $\btheta_k$ to generate $100$ data points for each device $k$. Our HM structure can be summarized as follows.
    \begin{align*}
    \bm{Y}_k|\btheta_k,\bm{\phi}&\sim\mathcal{N}(\bm{X}_k^\intercal\btheta_k,\sigma_k^2\bm{I})\\
    \btheta_k|\bm{\phi}&\sim\mathcal{N}(\bm{\mu},\text{diag}(\tau_1,\ldots,\tau_4))\\
    \bm{\mu}&\sim\mathcal{N}(\bm{0},\bm{I})\\
    \tau_i&\sim\log\mathcal{N}(0,1), \forall i\\
    \bm{\phi} &= (\bm{\mu},\log\tau_1,\ldots,\log\tau_4).
\end{align*}
    We calculate the posterior distribution of $\bm{\phi}$ using Algorithm \ref{algo:Fed-Gaussian} and plot the mean and $90\%$ CI for each component in Figure \ref{fig:CI} (Right). It can be seen that the mean of the posterior of $\bm{\phi}$ is close to the truth and the $90\%$ credible interval also covers the true parameter. This estimated $q(\bm{\phi})$ can be used as an initialization for new devices to achieve fast adaption.
    


\section{Real-World Case Studies}
\label{sec:exp}

\subsection{Student Performance Dataset}

This is a public dataset that can be found at \url{https://archive.ics.uci.edu/ml/index.php}. The dataset contains information on student performance (measured by exam scores) in secondary education of two Portuguese schools, namely, Gabriel Pereira and Mousinho da Silveira. It includes 29 predictors covering gender, grades, demographic and many other social/school-related features. Detailed information can be found in \cite{cortez2008using}. We treat each school as a ``device" (i.e., $K=2$). On each device, we randomly pick 60\% of the students as the training dataset and another 40\% of the students as the testing dataset. We create dummy variables for all nominal variables such as job and guardian. All other numeric variables are standardized to a zero mean and one standard deviation, following the guide in \cite{cortez2008using}. This data processing yields 38 predictors.

Our first goal is to select relevant predictors using our federated penalized regression technique (See Sec. \ref{sec:HM2-variable}). We then use the selected predictors to predict the final exam grades of students. We consider two most widely-used variable selection methods: Lasso and Ridge. Results are reported in Table \ref{table:school}. The model performance is evaluated based on the RMSE and number of included predictors. 

\begin{table}[!htbp]
\centering
\scriptsize
\begin{tabular}{cccc}
\hline
\textbf{Methods} & \textbf{A-RMSE} & \textbf{Number of included predictors (School 1)} & \textbf{Number of included predictors (School 2)}\\ \hline
Lasso (\texttt{HM2}) & 0.825 & 21 & 23 \\ \hline
Ridge (\texttt{HM2}) & 0.817 & 21 & 22 \\ \hline
\textbf{Methods} & \textbf{RMSE} & \multicolumn{2}{c}{\textbf{Number of included predictors}}\\ \hline
Lasso (Centralized)  & 0.820 &  \multicolumn{2}{c}{21} \\ \hline
Ridge (Centralized) & 0.815 &  \multicolumn{2}{c}{21} \\ \hline
\end{tabular}
\caption{The RMSE and number of included predictors for different federated variable selection methods.}
\label{table:school}
\end{table}

It can be seen that variable selection performance of \texttt{HM2} is consistent with the centralized variable selection method such as Lasso and Ridge regressions. This implies that our framework can serve as a new paradigm for decentralized variable selection problems. Additionally, \texttt{HM2} also yields comparable A-RMSEs compared to centralized methods. This demonstrates the advantage of borrowing strength from other devices. However, please note that, in terms of the prediction accuracy, federated variable selection can rarely beat the centralized approach as the latter one uses more data.


\subsection{NASA Aircraft Gas Turbine Engines}

In this case study, we consider condition monitoring data generated from aircraft gas turbine engines using the NASA commercial Modular Aero-Propulsion System Simulation (C-MAPSS) tools. The dataset is available at \url{https://ti.arc.nasa.gov/tech/dash/groups/pcoe/}. This dataset contains 100 engines. In each engine, 24 sensors are installed to collect time-series degradation signals. For each engine, we treat the first $60\%$ of the time-series observations as the training dataset and the remaining $40\%$ of the signals as the testing dataset. Within the training dataset, we sample $20\%$ of the data as a validation dataset. Our goal is therefore to predict the sensor signal trajectory on each gas turbine engine by training Algorithm \ref{algo:cor} using the training dataset. In this scenario, each engine can be viewed as a device (i.e., $K=100$).

It can be observed that all signal trajectories exhibit polynomial patterns and therefore many existing works resort to polynomial regression to analyze this dataset \citep{liu2013data, song2018statistical}. Here we detail the modeling procedure. Given a specific sensor, for all $k\in\{1,\ldots,K\}$, device $k$ fits a $d$-th order polynomial regression in the form of
\begin{align*}
    \bm{Y}_k = \bm{X}_k^\intercal\btheta_k + \text{noise},
\end{align*}
where the $(d+1)\times N_k$ design matrix $\bm{X}_k$ is in the form of
\begin{align*}
    \bm{X}_k^\intercal = 
    \begin{bmatrix}
    1 & [x_{k1}]_1 & [x_{k1}]_2^2 & \ldots & [x_{k1}]_d^{d}\\
    1 & [x_{k2}]_1 & [x_{k2}]_2^2 & \ldots & [x_{k2}]_d^{d}\\
    \vdots & \vdots & \vdots & \ldots & \vdots\\
    1 & [x_{kN_k}]_1 & [x_{kN_k}]_2^2 & \ldots & [x_{kN_k}]_d^{d}
    \end{bmatrix}.
\end{align*}
In the above expression, $\bm{Y}_k$ represents the signal trajectory for device $k$ and $\bm{x}_{k1}$ represents time. In \texttt{HM1}, device parameters $\{\btheta_k\}_k$ are estimated using Algorithm \ref{algo:cor}.

We benchmark our proposed model with the following algorithms:

\begin{itemize}
    \item \texttt{FedAvg}: FedAvg is one of the most fundamental and competing benchmark models in the FDA. During each communication round, device $k$, for all selected $k$, first runs several steps of local SGD and then sends updated parameters $\btheta_k$ back to the server. The server the aggregates those parameters by calculating $\bbtheta=\frac{1}{|\mathcal{S}|}\sum_{k\in\mathcal{S}}\btheta_k$. This step is repeated several times and ultimately each device will use the global parameter $\bbtheta$ to perform prediction.
    \item \texttt{Ditto}: \texttt{Ditto} is a personalized FL algorithm. The first stage of \texttt{Ditto} is the same as \texttt{FedAvg} and generates a global parameter $\bbtheta$. Afterwards, each device $k$ derives the personalized solution $\bm{v}_k$ by solving a constrained optimization problem $F_k(\bm{v}_k)+\frac{\lambda_{Ditto}}{2}\norm{\bm{v}_k-\bbtheta}_2^2$ where $F_k(\cdot)$ is the local loss function and $\lambda_{Ditto}$ is a tuning parameter. The intuition is that, each device can run several updating procedures to collect personalized solutions while this solution stays in the vicinity of the shared global model to retain useful information from a global model.
    \item \texttt{Separate}: In \texttt{Separate}, each device simply fits its own linear model without communication.
\end{itemize}

For all models, we set $T=20, C=100$, and use grid-search to tune the learning rate (and other model hyper-parameters). In Algorithm \ref{algo:cor}, we set $\alpha=0.9$. We report the A-RMSE across all $100$ devices in Table \ref{table:nasa}. 

\begin{table}[!htbp]
\centering
\begin{tabular}{ccccc}
\hline
 Sensor &  \texttt{HM1} ($\alpha=0.9$) & \texttt{Separate} & \texttt{FedAvg} & \texttt{Ditto} \\ \hline
 Sensor 2 &   $\bm{0.270(\pm0.001)}$ & $0.299(\pm 0.003)$ & $0.450(\pm 0.013)$ & $0.281(\pm 0.001)$\\ \hline
 Sensor 3 &  $\bm{0.218(\pm 0.002)}$ & $0.223(\pm 0.001)$ & $0.303(\pm 0.009)$ & $0.220(\pm 0.001)$ \\ \hline
 Sensor 7 &  $\bm{0.369(\pm 0.004)}$ & $0.405(\pm 0.003)$ & $0.628(\pm 0.011)$ & $0.388(\pm 0.005)$\\ \hline
 Sensor 8 &  $\bm{0.267(\pm 0.001)}$ & $0.307(\pm 0.001)$ & $0.395(\pm 0.008)$ & $0.289(\pm 0.001)$ \\ \hline
\end{tabular}
\caption{The A-RMSE of all model over 30 independent runs. We report the standard deviation in the brackets.}
\label{table:nasa}
\end{table}

From Table \ref{table:nasa}, it can be seen that \texttt{FedAvg} consistently yields the worst prediction accuracy as one shared global parameter $\bbtheta$ does not suit all devices, especially in a heterogeneous setting. Personalized approaches such as \texttt{Ditto} circumvent this disadvantage of global models and generate personalized solutions for each device. Those personalized methods, however, ignore related information amongst devices. Our method, on the other hand, improves the prediction accuracy by exploiting a joint structure for inductive transfer.

\section{Conclusion}
\label{sec:con}

In this paper, we propose a federated treatment for linear regression by adopting a hierarchical modeling approach. We test our proposed framework on a range of simulated and real-world datasets. Despite the simplicity of our linear model framework, it can outperform many state-of-the-art federated algorithms and we argue that it can serve as a competing benchmark model for future development of federated algorithms. One possible future direction is to extend our framework to the generalized linear models such as linear mixed-effect models or to more complicated models such as Gaussian processes and tensor regression. We hope our work will help inspire continued exploration into the world of federated data analytics and its engineering applications.

\bibliography{mybib}
\bibliographystyle{apalike}
\end{document}